\documentclass[useAMS,usenatbib]{mn2e}
\usepackage{graphicx}

\title{The first spectroscopic verification of an extragalactic classical
chemically peculiar star}
\author[Paunzen, Netopil \& Bord]
   {E.~Paunzen$^{1}$\thanks{E-mail: ernst.paunzen@univie.ac.at}
   \thanks{This paper includes data gathered with the 6.5 meter
   Magellan Telescopes located at Las Campanas Observatory, Chile.}, M.~Netopil$^{1}$, and 
   D.J.~Bord$^{2,3}$\\
  $^{1}$Institut f{\"u}r Astronomie der Universit{\"a}t Wien, 
  T{\"u}rkenschanzstr. 17, A-1180 Wien, Austria \\
  $^{2}$Department of Natural Sciences, University of Michigan-Dearborn, 
  Dearborn, MI 48128-1491, USA \\
  $^{3}$Department of Astronomy, University of Michigan, Ann Arbor, MI    48109-1042, USA}
\date{Accepted 2010 September 9. Received 2010 August 30; in original form 2010 July 27}

\pagerange{\pageref{firstpage}--\pageref{lastpage}} \pubyear{2010}

\def\LaTeX{L\kern-.36em\raise.3ex\hbox{a}\kern-.15em
    T\kern-.1667em\lower.7ex\hbox{E}\kern-.125emX}

\begin{document}

\label{firstpage}

\maketitle

\begin{abstract}
We present the first spectroscopic verification of a bona fide chemically peculiar (CP) star in the
Large Magellanic Cloud. CP stars reside on the upper main sequence and are characterized by
strong global stellar magnetic fields with a predominant dipole component oriented at random
with respect to the stellar rotation axis and displaced from the star's centre. Overabundances
with respect to the Sun for heavy elements such as silicon, chromium, strontium and europium
are also a common phenomenon. These objects are excellent astrophysical laboratories by
which to investigate many of the processes connected with star formation and evolution.
Several studies comparing the incidence of CP stars in the Large Magellanic Cloud with that
of the Milky Way have been published. These investigations are based on the photometric
detection of CP stars via the $\Delta a$ system which has been tested and calibrated for objects in the
Milky Way. From our spectroscopic observations made at Las Campanas Observatory, we are
able to confirm one classical B8\,Si star among the photometric sample, as well as one early
B-type emission-line star which was also initially detected by its significantly deviating $\Delta a$
value. We conclude that classical extragalactic CP stars do exist and that the photometric $\Delta a$
system is able to detect them in an efficient way.
\end{abstract}


\section{Introduction}

The group of chemically peculiar (CP) stars of the upper main sequence display
metal abundances that deviate significantly from the standard abundance distribution
\citep{As2009}; a subset of this class, the CP2 stars \citep{Pr1974}, reveals 
the existence of strong global stellar magnetic fields \citep{Co2009}. 
In addition, this subclass of B to F-type stars is characterized 
by radial velocity changes as well as photometric variability due to
stellar rotation and a spotty surface structure. 

The origin of the stellar magnetic fields is still not resolved. 
Several recent publications (e.g. \citealt{Hu2009})
favour a model in which they are the survivors of frozen-in fossil fields
originating from the medium out of which the stars were
formed. During the main sequence evolution, a dynamo
mechanism may act in the interior of these stars, enhancing
the field strength.

One of the most important key parameters for our understanding of 
star formation and evolution is the intrinsic metallicity of (proto-)stars 
of a given mass. Even in the early stages of stellar evolution, the metallicity 
severely influences the cooling and collapse of ionized gas \citep{Ja2007}.
With this fact in mind, by investigating CP stars in the Magellanic 
Clouds, we are able to assess the degree to which the presence or
nature of chemical peculiarities is governed or influenced by initial metallicity.
Moreover, by such studies we may be in a position to determine
whether different magnetic field strengths in 
the region of star formation lead to the same frequency of magnetic stars.
The incidence and characteristics of CP2 stars are already well 
investigated in the Milky Way \citep{Ne2007}. 

Due to the typical flux depression in CP stars at $\lambda$~5200\AA, the tool 
of $\Delta a$ photometry is able to detect them in an economical and efficient 
way by comparing the flux at the centre (5200\AA, $g_2$) to the adjacent 
regions (5000\AA, $g_1$ and 5500\AA, $y$). 
It has been shown that virtually all peculiar stars with magnetic fields have significant
positive $\Delta a$ values of up to +100 mmag whereas Be/Ae and metal-weak stars exhibit 
significantly negative ones \citep{Pa2005b}. 

Since the first detection of photometric CP candidates by \citet{Ma2001} in the
Large Magellanic Cloud (LMC), several
other studies, summarized in \citet{Pa2005a}, significantly improved the observational
evidence that the incidence of these stars is much less than in the Milky Way.

However, all these investigations are based on one important assumption:
classical CP stars do exist in the LMC and are unambiguously detected by 
$\Delta a$ photometry.

In this paper, we present the first spectroscopic verification of one
bona-fide photometric CP star candidate in the LMC on the basis of the
strong \hbox{Si\,{\sc ii}} doublet at 4128/4131\AA. In addition, we
are able to confirm the emission feature of one object found to
exhibit a significant negative $\Delta a$ value. These results compare
well with those of Galactic investigations of the same kind \citep{Pa2005b}.

\begin{figure}
\begin{center}
\includegraphics[width=80mm]{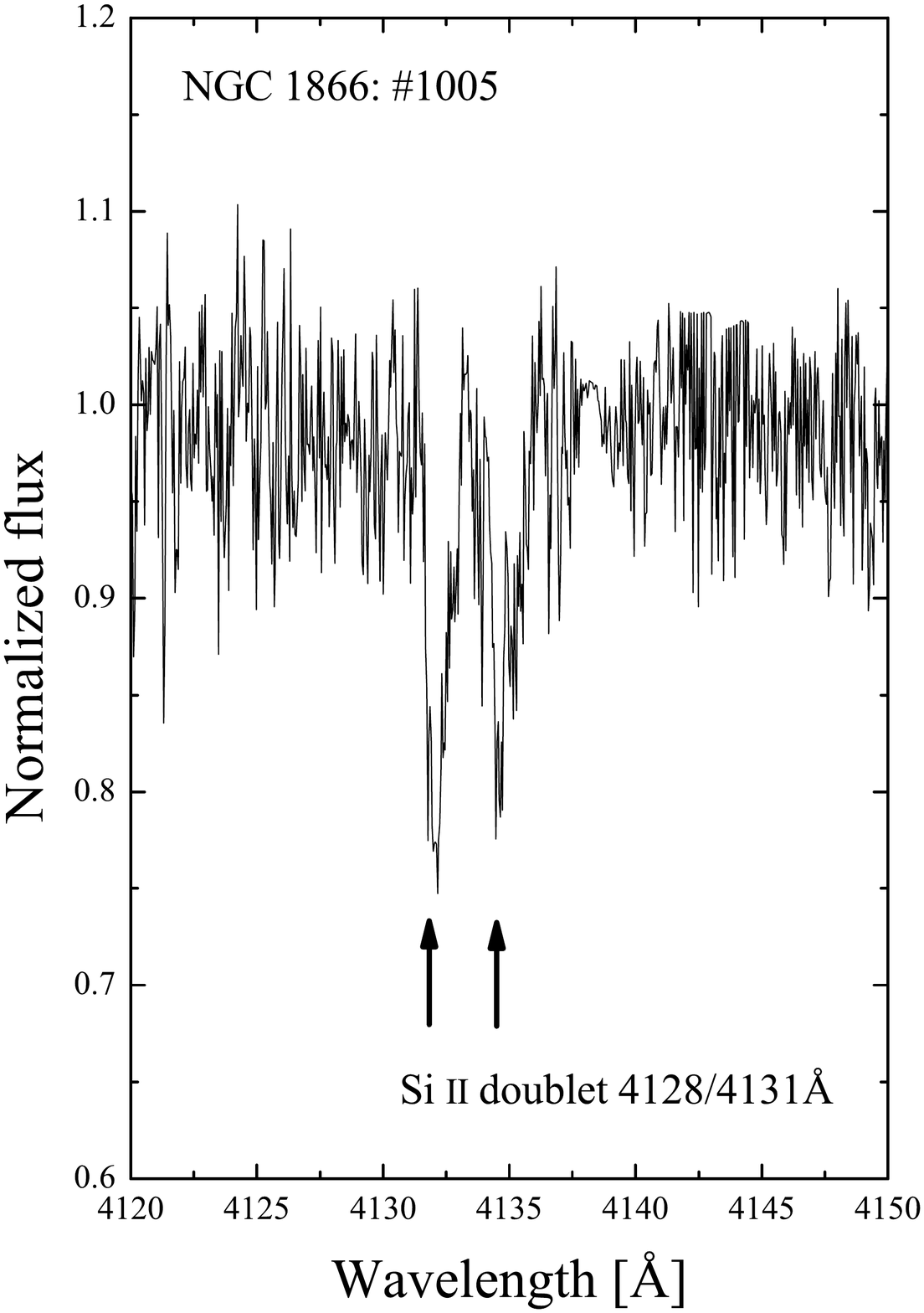}
\caption{The \hbox{Si\,{\sc ii}} doublet detected in star \#1005. 
The spectrum has been shifted to the rest frame of the Sun.  
To within the measurement uncertainties ($\sim$15\%), the equivalent 
width of each line is the same, viz. 0.30\,\AA.}
\label{plot_si}
\end{center}
\end{figure}

\begin{figure}
\begin{center}
\includegraphics[width=80mm]{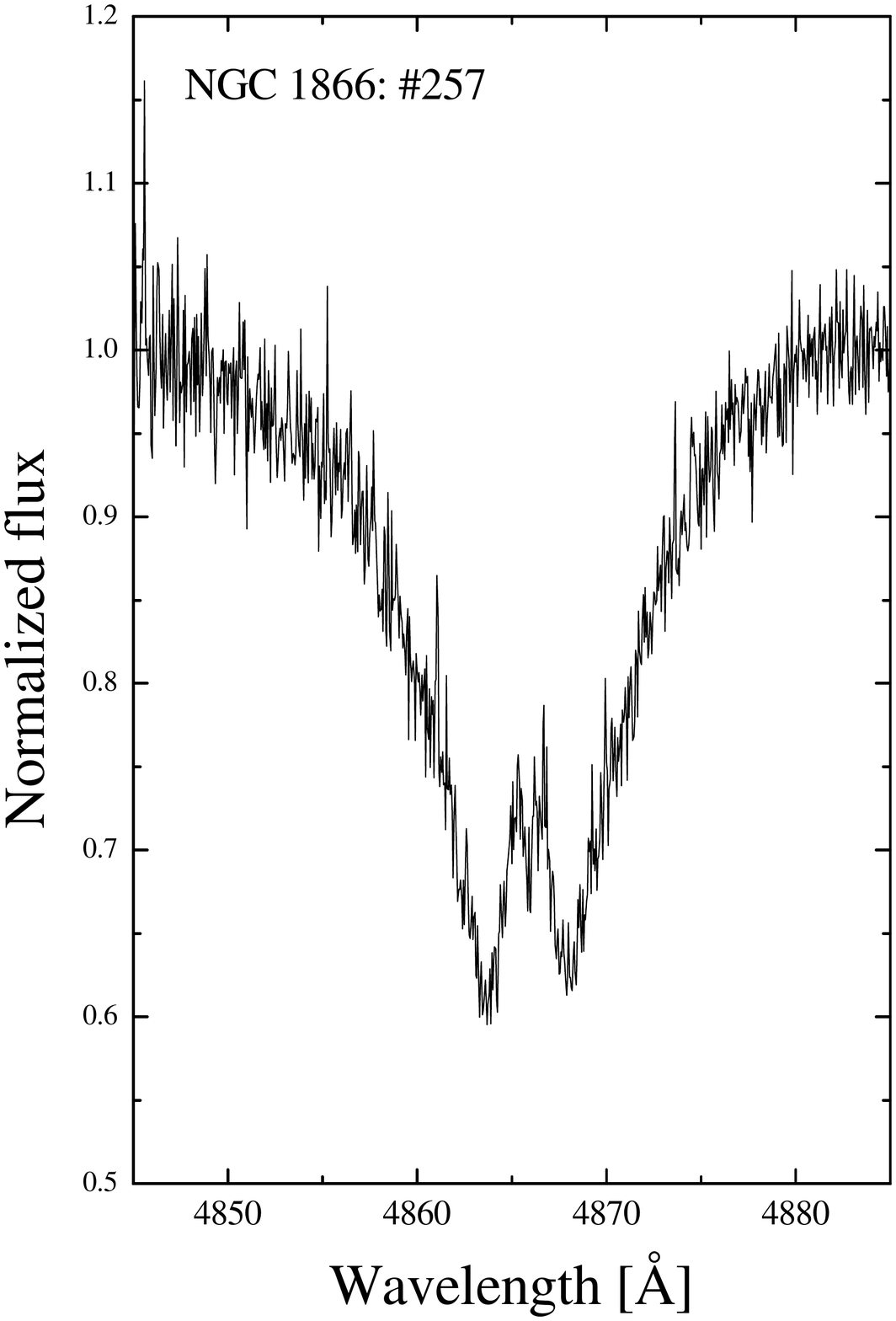}
\caption{The emission structure of H$\beta$ in star \#257 which
exhibits a significant negative $\Delta a$ value.
The spectrum has been shifted to the rest frame of the Sun.}
\label{plot_emission}
\end{center}
\end{figure}

\section{Target selection, observations and reductions}

From the published list of stars given by \citet{Ma2001} 
with significant positive $\Delta a$ values, we chose the
CP candidates \#687, \#1005, and \#1093 (the numbering system is adopted
from the aforementioned reference).
In addition, the metal-weak/Ae/Be/shell candidate \#257
was selected for our study. The $V$ magnitudes of the targets, all located
in the field of NGC~1866, are between 16.9 and 18.2\,mag, near the limiting
brightness for achieving usable spectra in moderate exposure times with the
instrumentation available.

The observations were performed with the Magellan Inamori Kyocera Echelle (MIKE), 
a high-throughput, double echelle spectrograph \citep{Be2003} on the 6.5-m
Magellan telescope (Clay) at Las Campanas (Chile) on the nights of
18-20 November 2004 by one of us (DJB). 
The spectrograph employs a dichroic filter with separate
blue and red fibres to provide simultaneous coverage of wavelengths spanning
the range from 3200 to
10\,000\AA\,with a typical one-pixel resolution of 0.04 to 0.05\AA.
The spectra were binned 2x2 to produce a resolution of about 0.08\,\AA\,in the blue. 
This value comports well with the measured half-widths of sharp, unblended 
comparison lines in the ThAr spectrum near 4000\,\AA\,which yields on the 
average 0.12\,\AA.
Integration times between 30 and 60 minutes were chosen
to control the intensity of the night sky lines and to limit spectral contamination
from neighbouring stars in crowded fields caused by rotation of the slit on the sky.

All reductions were performed with standard IRAF
routines\footnote{http://iraf.noao.edu/}. After the bias, dark and
flat field corrections, the spectra were wavelength calibrated (including
a heliocentric correction) using two ThAr standard 
spectra observed before and after the corresponding target.
As a final step, the spectra were co-added to increase the signal-to-noise ratio.

\section{Results and conclusions}

Due to the faintness and the high spectral resolution of the
instrument, the signal-to-noise ratios of the final spectra
are very low, typically around 15. For the bluest and
reddest regions of the spectra, no signal at all was 
observed because of the drop of the quantum efficiency 
of the CCDs.

For all stars, the hydrogen lines are the most prominent
features in the spectra. The photometric candidates are thus clearly
in the expected effective temperature range for classical
CP stars and not contaminated by any late-type fore- or background
objects. 

One of the most prominent features of early-type CP stars is the 
\hbox{Si\,{\sc ii}} doublet at 4128/4131\AA\, which is very
much stronger than comparable lines in normal stars \citep{Bi1955}.
The enhancement of this doublet is an unambiguous characteristic
of magnetic CP stars in general, and so we primarily searched for this
feature in the spectra. 

For targets \#687 and \#1093 we were not able to detect any features 
beside the hydrogen lines due to the low signal-to-noise ratio of the spectra 
and moderate rotational velocities of the stars.  Applying several smoothing 
techniques (boxcar, FFT, and median) still revealed no additional features 
above 5$\sigma$ of the mean noise level.  Unfortunately, we are not able to 
reliably analyze the 5200\AA\, region because the depths of the flux 
depressions for the measured $\Delta a$ values for these stars are too shallow 
for the low signal-to-noise ratio \citep{Ku2004}.

The rotational speeds of these two stars can be estimated, however, 
from comparisons of the observed hydrogen line profiles with synthetic 
spectra. Establishing a credible continuum level for the hydrogen 
lines in our echelle spectra is neither easy nor unambiguous, and 
normalization was carried out using the shape of the flat field 
function of the corresponding aperture.  From different numerical 
experiments using a range of stellar parameters, we find that the most probable
$v$\,sin\,$i$ values for \#687 and \#1093 fall within the range of 100 to 150
km\,s$^{-1}$.

Figure \ref{plot_si} shows the \hbox{Si\,{\sc ii}} doublet observed
for \#1005 with the most significant $\Delta a$ value detected, 
10.8$\sigma$ above the normality line. We measured the separation
of the two lines which is, within the errors, identical with the
theoretical one. The spectral feature is
so strong that we immediately classify this object as a typical
silicon star of the CP2 group. 
We measured the equivalent widths of the doublet to be
0.30\,\AA\, for each line with an error of about 15\%. For normal
late B-type stars (dwarfs and giants), the corresponding value is 
0.19\,\AA\, \citep{Ja1995}. We notice that the equivalent width of the
\hbox{Si\,{\sc ii}} doublet for dwarfs is at a maximum for
spectral type B8 and drops to 0.15 and 0.11\,\AA\, for B7 and
B9, respectively. The corresponding equivalent widths for late B-type CP stars range 
from 0.2 to 0.4\,\AA, depending on the rotational phase and the strength of the
stellar magnetic field \citep{Lo1999}. These values are in agreement with the
measurements of \#1005.

\citet{Te1999} published
$V$\,=\,17.797(5) and $(B-V)$\,=\,0.014(13)\,mag for this star.
Using the distance modulus of 18.33  
and $A_{\rmn{V}}$\,=\,0.25\,mag for NGC~1866 and its surrounding
taken from \citet{Sa2003}, we derive an absolute magnitude of
$-$0.8\,mag which coincides with a main sequence late B-type star.
Within the given errors we classify \#1005 as B8\,Si star.
From the original as well as a smoothed spectrum, we determined the upper limit
of $v$\,sin\,$i$ as 50 km\,s$^{-1}$ from the hydrogen lines and 
\hbox{Si\,{\sc ii}} doublet.

The emission characteristics of H$\beta$ for star \#257, which
exhibits a significant negative $\Delta a$ value at a 10.7$\sigma$
significance level, is shown in Fig. \ref{plot_emission}. 
This object is about 0.7\,mag brighter that \#1005 which suggests
an early B-type. Again, this result is in agreement with the findings
for similar stars in the Milky Way.

From the wavelength calibrated spectra (Figs. \ref{plot_si} and 
\ref{plot_emission}), using the hydrogen lines, we
were also able to obtain estimates of the heliocentric radial velocities
for each target star. The final values are between 280 and 320 km\,s$^{-1}$
with errors of about 20 km\,s$^{-1}$. This clearly establishes
them as true members of the LMC and not Galactic foreground objects
\citep{Pr1985}.

With the first spectroscopic verifications of the true natures
of photometrically detected peculiar stars in the LMC, we demonstrate the reliability
of the $\Delta a$ system for observations in extragalactic environments. 
The published photometric CP candidates are therefore excellent
targets for follow-up investigations which will help shed more
light on stellar formation and evolution in areas with different
metallicities and/or different magnetic field strengths.
For the latter, no measurements in the Magellanic Clouds are yet available.
With new generation telescopes
and satellite missions, we anticipate that more distant galaxies in and beyond the 
Local Group will become accessible for this research area.

\section*{Acknowledgements}
The authorswish to thank the referee, Dr George Preston, for several
helpful comments that led to improvements in this paper. This work
was supported by the financial contributions of the Austrian Agency
for International Cooperation in Education and Research (WTZ CZ-
10/2010 and HR-14/2010). MN acknowledges the support by the
O¨ FG under MOEL grant nos 388 and 446.

\label{lastpage}

\end{document}